\documentclass[manuscript]{emulateapj}

\newcommand{\targ}{\object{1E1740.7-2942}}
\newcommand{\nus}{{\em NuSTAR}}
\newcommand{\suz}{{\em Suzaku}}
\newcommand{\xte}{{\em RXTE}}
\newcommand{\intg}{{\em INTEGRAL}}

\slugcomment{Accepted by the Astrophysical Journal}

\shorttitle{{\emph NuSTAR} and {\emph INTEGRAL} observations of 1E1740.7-2942}
\shortauthors{Natalucci et al.}

\begin{document}


\title{\nus~ and \intg~ observations of a low/hard state of 1E1740.7-2942}


\author{Lorenzo~Natalucci\altaffilmark{1}, John~A.~Tomsick\altaffilmark{2},
        Angela~Bazzano\altaffilmark{1}, David~M.~Smith\altaffilmark{3},
        Matteo Bachetti\altaffilmark{4,5}, Didier~Barret\altaffilmark{4,5}, 
        Steven~E.~Boggs\altaffilmark{2}, Finn~E.~Christensen\altaffilmark{6},
        William~W.~Craig\altaffilmark{2,7}, Mariateresa~Fiocchi\altaffilmark{1},
        Felix~F\"urst\altaffilmark{8}, Brian~W.~Grefenstette\altaffilmark{8}, 
        Charles~J.~Hailey\altaffilmark{9}, Fiona~A.~Harrison\altaffilmark{8},
        Roman~Krivonos\altaffilmark{2}, Erik~Kuulkers\altaffilmark{10},
        Jon~M.~Miller\altaffilmark{11}, Katja Pottschmidt\altaffilmark{12,13},
        Daniel~Stern\altaffilmark{14}, Pietro~Ubertini\altaffilmark{1},
        Dominic~J.~Walton\altaffilmark{8}, William~W.~Zhang\altaffilmark{15}}
        
\altaffiltext{1}{Istituto Nazionale di Astrofisica, INAF-IAPS, via del Fosso del Cavaliere,
00133 Roma, Italy; e-mail: lorenzo.natalucci@iaps.inaf.it}
\altaffiltext{2}{Space Sciences Laboratory, 7 Gauss Way, University of California, Berkeley, 
CA~94720--7450, USA}
\altaffiltext{3}{Physics Department and Santa Cruz Institute for Particle Physics, 
University of California, Santa Cruz, CA~95064, USA}
\altaffiltext{4}{Universit\'e de Toulouse; UPS-OMP; IRAP; Toulouse, France}
\altaffiltext{5}{CNRS; Institut de Recherche en Astrophysique et Plan\'etologie; 
9 Av. colonel Roche, BP 44346, F-31028 Toulouse cedex 4, France}
\altaffiltext{6}{DTU Space, National Space Institute, Technical University of Denmark, 
Elektrovej 327, DK-2800 Lyngby, Denmark}
\altaffiltext{7}{Lawrence Livermore National Laboratory, Livermore, CA 94550, USA}
\altaffiltext{8}{Cahill Center for Astronomy and Astrophysics, California Institute of 
Technology, Pasadena, CA 91125, USA}
\altaffiltext{9}{Columbia Astrophysics Laboratory, Columbia University, New York, NY 10027, USA}
\altaffiltext{10}{European Space Astronomy Centre (ESA/ESAC), Science Operations Department, 
28691 Villanueva de la Ca\~{n}ada (Madrid), Spain}
\altaffiltext{11}{Department of Astronomy, University of Michigan, 500 Church Street,
Ann Arbor, MI~48109, USA}
\altaffiltext{12}{CRESST and NASA Goddard Space Flight Center, Astrophysics Science
Division, Code 661, Greenbelt, MD 20771, USA}
\altaffiltext{13}{Center for Space Science and Technology, University of Maryland Baltimore
County, 1000 Hilltop Circle, Baltimore, MD 21250, USA}
\altaffiltext{14}{Jet Propulsion Laboratory, California Institute of Technology, 
Pasadena, CA 91109, USA}
\altaffiltext{15}{NASA Goddard Space Flight Center, Greenbelt, MD 20771, USA}

\begin{abstract}
The microquasar \targ, also known as the ``Great Annihilator'', was observed by \nus~ in the 
Summer of 2012. We have analyzed in detail two observations taken $\sim2$~weeks apart, for which 
we measure hard and smooth spectra typical of the low/hard state. A few weeks later the source 
flux declined significantly. Nearly simultaneous coverage by \intg~is available from 
its Galactic Center monitoring campaign lasting $\sim2.5$~months. These data probe the hard state 
spectrum from \targ~before the flux decline. We find good agreement between 
the spectra taken with IBIS/ISGRI and \nus, with the measurements being compatible with 
a change in flux with no spectral variability. We present a detailed analysis of the 
\nus~spectral and timing data and upper limits for reflection of the high energy emission. 
We show that the high energy spectrum of this X-ray binary is well described by 
thermal Comptonization. 
\end{abstract}


\section{Introduction}
The astrophysical source \targ~is a known microquasar located near the 
Galactic Center (GC), at an angular distance of 50$^\prime$ from SgrA$^{*}$. 
First discovered by {\em Einstein}/IPC \citep{her84} in
the soft X-rays, it is the most luminous persistent source above $20$~keV in the 
region \citep{sun91}, and has extended radio lobes reaching distances of up to a few parsecs 
($\sim1^\prime$) from its core \citep{mir92}. The core radio emission is found to be variable,
with radio flux and spectral slope changes that are correlated with the X-ray flux, as observed
by {\em GRANAT}/SIGMA in the early 1990's \citep{pau91}. SIGMA reported a burst 
of emission in soft $\gamma$-rays characterized by a broad hump
in the $300-600$~keV band \citep{bou91} and a further, more marginal episode of enhanced 
$\gamma$-ray emission \citep{chu93}. It was then speculated that such transient 
events could be generated by the 
deceleration and interaction of positrons injected by the source into a 
molecular cloud \citep{bal91,mir93}. However, nearly simultaneous observations by {\em CGRO}, 
namely by OSSE \citep{jun95} and BATSE \citep{smi96}, did not detect any 
transient emission and also, the high energy observations
by \intg~\citep{bou09} and other satellites could not
confirm the high energy feature reported by SIGMA.

Many aspects about the nature of \targ~ remain a mystery. 
A clear optical/IR counterpart for the companion star has not been detected so far, 
probably due to the source environment chacterized by a high concentration of dust and a 
large hydrogen column density (${N}_{\rm H}\sim10^{23}$~cm$^{-2}$). Therefore, its nature 
as a high mass or low mass object is not known, nor its distance and inclination. However,
the high amount of absorption and its position near the GC favors a distance
of $\sim8.5$~kpc, and the presence of bipolar jets disfavors a face-on geometry for the 
accretion disk. Recently, \citet{mar10} reported a candidate companion 
consisting of a single near-infrared source with an apparent non-stellar morphology, 
localized at a position coincident with the source radio core. 

Despite its brightness, so far there is no evidence for strong reprocessing from \targ.
The \suz~data established the absence (or evidence of weakness) 
of a reflection component in the hard X-ray spectra and the weakness of the iron line
\citep{rey10,nak10}.

There are remarkable similarities between the spectral and timing characteristics of \targ~and
those of black hole candidates (BHC) like \object{Cyg X-1}. For both sources, \citet{kuz97} 
found a positive correlation between spectral hardness and hard X-ray luminosity for 
${L_{\rm x}}$~$\lesssim10^{37}$~erg/s. Similarly, the 14~year-long \xte~ monitoring observed a
hysteresis effect relative to the hard state indicating an anticorrelation of the 
power law (PL) index with the derivative of the photon flux (i.e., the spectrum 
softens for decreasing flux). \citet{smi02} also reported that in a single \xte~ pointing,
the spectral index hardened dramatically while the total flux of photons remained unchanged.
The \xte~ monitoring data was also used to search for periodicities of \targ~after filtering out 
the long-term variations. This resulted in the detection of a 
narrow signal at 12.7~days in the Lomb-Scargle periodogram. If this is the orbital period and
if the binary system accretes by Roche-lobe overflow, the companion should be a red giant. 
A much longer periodicity of $\sim600$~days has also been reported, possibly related to 
cyclic transitions
between a flat and warped disk as observed in other luminous binaries like 
\object{Cyg X-1} and \object{LMC X-3} \citep{ogi01}.    

\begin{figure}
\centering
\hspace{1cm}
\includegraphics[angle=90,scale=.40]{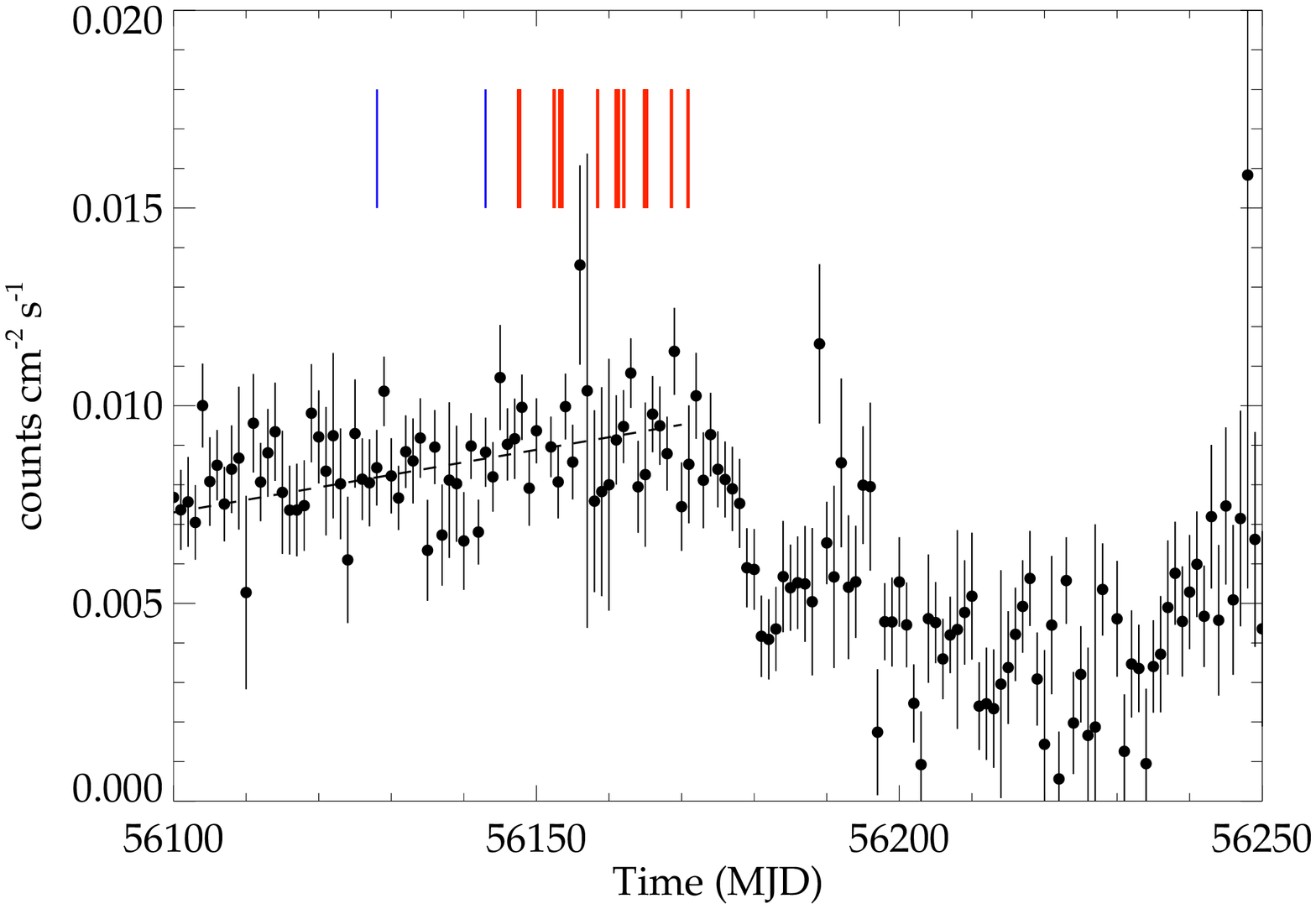}
\caption{The {\em Swift}/BAT light curve (LC) of 1E1740.7--2942~in the 15--50~keV energy band. The
\nus~and \intg~observations are shown as vertical lines in blue and red,
respectively. During the overall time period of the observations the source is 
bright at a level of $\sim35$~mCrab. The dashed line shows 
the section of the LC before the flux decline, fitted by a linear function.}
\label{fig_batlc}
\end{figure}

\begin{table*}
\begin{center}
\caption{\nus~ and \intg~ observations of 1E1740.7--2942. \label{table_observations} }
\begin{tabular}{lcccccc}
\tableline\\ [-2.0ex]
Obs.id & Epoch & Revolution & Satellite & Start (UTC) & End (UTC) & Exposure Time (s) \\ 
\tableline\\ [-2.0ex]
10002021001 & 1 & -            & \nus  & 2012-07-19 23:36:07 & 2012-07-20 02:11:07 & 2172  \\
10002023003 & 2 & -            & \nus  & 2012-08-03 22:16:07 & 2012-08-04 01:06:07 & 6125  \\
09200250033 & 3 & { 1199 } & \intg & 2012-08-08 10:16:45 & 2012-08-08 19:08:37 & 15861 \\
09200010015 & 3 & { 1200 } & \intg & 2012-08-13 08:45:09 & 2012-08-13 12:56:12 & 9615 \\
09200250007 & 3 & { 1201 } & \intg & 2012-08-14 03:00:31 & 2012-08-14 14:48:53 & 22472 \\
09200010016 & 3 & { 1202 } & \intg & 2012-08-19 08:16:17  & 2012-08-19 12:31:43 & 10030 \\
09200250003 & 3 & { 1203 } & \intg & 2012-08-21 21:33:04 & 2012-08-22 09:49:30 & 21614 \\
09200010017 & 3 & { 1204 } & \intg & 2012-08-22 23:19:34 & 2012-08-23 03:15:31 & 8757 \\
09200250004 & 3 & { 1205 } & \intg & 2012-08-25 19:16:27 & 2012-08-26 07:05:14 & 19728 \\
09200010018 & 3 & { 1206 } & \intg & 2012-08-29 12:42:07 & 2012-08-29 16:36:44 & 8635 \\
09200250008 & 3 & { 1207 } & \intg & 2012-08-31 18:47:36 & 2012-09-01 00:05:58 & 10219 \\
\tableline\\ [-2.0ex]
\end{tabular}
\tablecomments{Exposure times are the live exposure time of spectra for \nus~and IBIS/ISGRI.  
} 
\end{center}
\end{table*}

The spectrum of \targ~in the hard state is usually well fitted by a cutoff power-law (CPL) or 
Comptonization model such as {\it comptt} \citep{tit94,hua95}. \citet{bou09} reported the 
existence of a high energy excess dominating beyond $\sim200$~keV, 
which is reminiscent of a similar component observed for Cyg~X-1 \citep{mcc02}. 
\citet{jou12} presented a sensitive spectrum for Cyg~X-1 up to a few hundred keV with 
the SPI instrument \citep{ved03} on board \intg~\citep{win03}. They 
detected a high level of polarization for energies $>150$~keV;    
polarized emission was also
reported from the IBIS instrument \citep{lau11}. These observations support a jet origin of the excess
for Cyg~X-1. The SPI observations of
\targ~ are contaminated by the GC environment, especially at the lowest energies
(E$<100$~keV), due to the wide angular resolution of that instrument ($2.6^{\circ}$~FWHM). 
The SPI spectrum measured 
by \citet{bou09} was obtained by subtracting the contaminating source spectra as measured 
by IBIS/ISGRI. Conversely, the IBIS instrument \citep{ube03} allows unambiguous measurements 
from 20~keV up to a few hundred keV. 
\citet{rey10} reported on two observations separated by about 700 days with \suz. In
both cases, the source was observed after a transition to the hard state.  Significant emission 
was detected up to 300 keV showing no spectral break. The spectra were consistent with a Comptonized 
corona with the high emission accounted for with a hard power law ($\Gamma\sim1.8$) and a 
significant contribution from an accretion disk with $kT\approx0.4$~keV at soft X-ray energies. These 
authors reported that the accretion disc is not truncated at large radii, while 
residing close to the inner-most stable circular orbit (ISCO), 
i.e.   $R_{\rm in}\sim20R_{\rm g}$.  There is no significant detection 
of disc reflection in the \suz~data while a weak, broad iron line was marginally 
found at $E\sim6.7$~keV. 
Simultaneous observations of \targ~by \xte~and \intg/IBIS, obtained mostly when the source 
was in the hard state, were analyzed by \citet{del05} who reported values of the reflection 
normalization in the range $\sim0.3-0.9$. 
We note that their analysis is so far the only 
claimed detection of reflection in this source. 

The recently launched \nus~ 
observatory \citep{har13} is the ideal instrument for the study of the reprocessed 
components in the spectrum of \targ, due its unprecedented broadband coverage from soft
to hard X-rays. In this work, we use nearly simultaneous observations by \nus~ 
and the IBIS/ISGRI instrument on board \intg. 
IBIS, with its wide FOV and good sensitivity
near 100~keV, can be used to complement the \nus~ energy coverage and
better constrain the parameters of the Comptonization. 
For \intg~we limited our analysis to the IBIS data and did not attempt to use data from either 
Jem-X \citep{lun03} or SPI, mainly because Jem-X had a relatively short exposure time and the SPI 
data are seriously contaminated by other sources. We also searched for nearly simultaneous 
observations of \targ~in the {\em Swift}/XRT archive, but only very short, off-axis exposures 
were available. We verified that the statistics are too low to discriminate 
between spectral models from these data.

In Section \ref{sect_obs}, we describe the observations; Section \ref{analysis_nustar} deals
with the spectral and timing analysis of the \nus~data while in Section \ref{analysis_integral}
we present the IBIS/ISGRI spectrum and lightcurves. Section \ref{analysis_all} reports on the
joint \nus/\intg~spectrum. In Section \ref{sect_disc}, our results are compared to previous
observations of \targ, and finally our conclusions are presented in Section \ref{sect_conc}.
  
\section{Observations} \label{sect_obs}
All our observations were taken from July to September 2012. Data from \nus~are 
available from two epochs in July-August 2012, spaced by about two weeks 
(see details in Table \ref{table_observations}). The total exposure times were
2172s and 6125s, respectively. \nus~is the first X-ray satellite with  
multilayer hard X-ray optics and is operational in the energy range 
3--79 keV \citep{har13}. 
The mission carries two telescopes with grazing incidence optics, each one 
focusing on separate detector modules at a distance of 10m, i.e. two detectors named Focal 
Plane Modules A and B (FPMA, FPMB). These CdZnTe detectors have 
a spatial resolution of 0.6mm  and sample a total Field-Of-View (FOV) of $13^{\prime}$. 
The telescope Point Spread Function (PSF) has an $18^{\prime\prime}$ 
Full-Width-At-Half-Maximum with 
extended tails resulting in a Half-Power-Diameter of  $58^{\prime\prime}$.

Shortly after the \nus~observations were completed, \intg~began a  
GC monitoring campaign as part of its AO10 cycle. Furthermore, some of
the Galactic Bulge Monitoring data were available.
Due to the wide FOV ($29^{\circ}$~Full-Width-At-Zero Response) of
IBIS/ISGRI, \targ~ was monitored over quite a long period. Figure  
\ref{fig_batlc} shows the {\em Swift}/BAT light curve of \targ~taken around the time of 
our observations. The source experienced a significant flux decline near MJD~56175 
followed by a decrease of the accretion rate that was further monitored 
by \intg. For the purpose of this 
work, we selected data in a time period comprised between the 
first \nus~ observation and the latest date available before the flux decline.   
This results in a time coverage spanning nine \intg~orbits from 2012 August 8 to 
September 1 (revolutions 1199-1207).
These observations consist of a series of pointings, selected for the source being
at an offset $<9^{\circ}$ from the IBIS instrument axis. 
Due to the high total flux of sources from the GC region, the sensitivity of
IBIS is reduced particularly in the lowest energy channels. 

The \nus~observations are not simultaneous with \intg~and on the basis of the light curve 
shown in Figure \ref{fig_batlc}, are expected to have a $\sim10$\% lower flux. However,
rate fluctuations up to $\sim20-30$\% or more are observed at daily time scale. We distinguish
three epochs consisting of the two \nus~observations and of the further \intg~monitoring. 
Since \targ~ does not usually exhibit substantial spectral variations within 
such a small flux range, we model the data from both satellites simultaneously,
allowing the cross-normalization to vary. 

Note that the total \nus~observation time is much shorter than the \intg~one. 
IBIS/ISGRI, being particularly sensitive in the $\sim50-100$~keV spectral region, 
is expected to provide a good overlap of both data sets, allowing minimal possible 
bias in the spectral modeling.         
\section{Data Analysis} \label{sect_analysis}

\subsection{\nus} \label{analysis_nustar}

\begin{figure}
\includegraphics[angle=0,scale=.45]{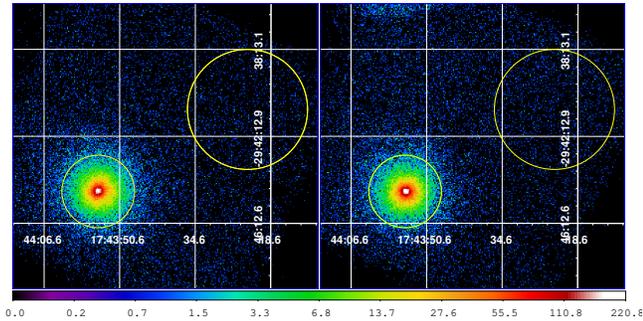}
\caption{Images of \nus~focal planes (FPMA and FPMB, from left to right) for the
epoch~2 observation. The images are in log-scale and the intensity unit is  
counts/pixel. Source and background extraction
regions are denoted by circles. \label{fig_images}}
\end{figure}

We analyzed the \nus~data using the \nus~Data Analysis Software ({\it NuSTARDAS})  
version 1.1.1, CALDB version 20130509 and in-flight calibrated response
matrices. The software applies offset correction factors to the energy 
response to account for the movement of the mast, causing a varying position of the 
focal spots on the detector planes. For the two FPMs the pipeline 
produces images, spectra and 
deadtime corrected lightcurves. For each \nus~observation, the source and background 
subtraction regions must be carefully evaluated due to the possible presence of contaminating
sources outside the FOV. This straylight problem can be minimized prior to the
observations by a tuning of the spacecraft position angle (PA) and if needed, of the 
optical axis position. In Figure \ref{fig_images} the images obtained for 
both FPMs are shown. The intensity is encoded in logarithmic scale, to emphasize
underlying structures. 
The source net spectra were obtained by selecting counts in a circular region of $90^{\prime\prime}$ radius 
centered on the source position and subtracting count rates measured in a background dominated 
region, also circular, of radius $183^{\prime\prime}$. The different area normalizations were 
taken into account in the background subtraction. Both regions are shown in the figure. 

The total source count rates in the energy range 3--75 keV, in the selected spatial region 
are: $8.66\pm0.06$~c/s (FPMA)
and $8.05\pm0.06$~c/s (FPMB) for epoch~1, and 
$8.13\pm0.04$~c/s (FPMA) and $8.12\pm0.04$~c/s (FPMB) for epoch~2.  
The ratio of the source-to-background rates varies with energy and is as high as  
$\approx25$ and 35 for the energy bands 3--10 keV and 10--40 keV, respectively. 
We checked against possible systematic effects introduced by spatial variations 
of the background rates. For the epoch~2 observation, we extracted a set of source 
spectra in different annular regions of the detectors, far from the source, 
and having the same size as the source region shown in Figure \ref{fig_images}. 
We found that the maximum
variation of the 3--60~keV rates in these spectra is $\approx10^{-2}$ counts 
s$^{-1}$, i.e. a fraction of only $\sim10^{-3}$ of the total source intensity.\\ 

{\em Spectral analysis ---} 
We first analyzed spectra of the two observations using XSPEC \citep{arn96} 
version 12.8.0 and models consisting of a PL and a PL with a high energy cutoff. 
To model the effect of X-ray absorption,
we used the XSPEC {\it TBabs} model with abundances set as in \citet{wil00} and
cross sections set as in  \citet{ver06}. 
Note that the values of the absorption column derived from these abundances may be 
different to previously reported determinations of $N_{\rm H}$. 
For each observation we fitted the two spectra obtained from FPMA and FPMB, 
rebinned to have at least 75 counts per spectral bin. This choice is appropriate 
for a good overlap with the IBIS spectra, i.e. it allows similar error sizes for the 
spectral bins in the range $\sim30-70$~keV.   
Since the absorption is quite
high, in all the fits we left the {\it TBabs} absorption column free.

The results are reported in Table \ref{table_spec_nuonly}. 
The spectra from both observations do not show clear evidence for a high energy cutoff,
though the $\chi^{2}_{\rm red}$ improves slightly for epoch~2 using the CPL model.
For this observation, 
a high energy cutoff is formally constrained as $141 < E_{\rm fold} <447$~keV 
(90\% confidence). An attempt to model a high energy cutoff for epoch~1 did not lead
to meaningful constraints. Furthermore, for epoch~2, if we require the 
exponential cutoff to occur only at energies above a certain threshold $E_{\rm cut}$ 
({\it powerlaw*highecut} for XSPEC), the latter parameter is compatible with zero 
or, formally, $E_{\rm cut}<14$~keV at 90\% confidence.

\begin{figure}
\includegraphics[angle=270,scale=.32]{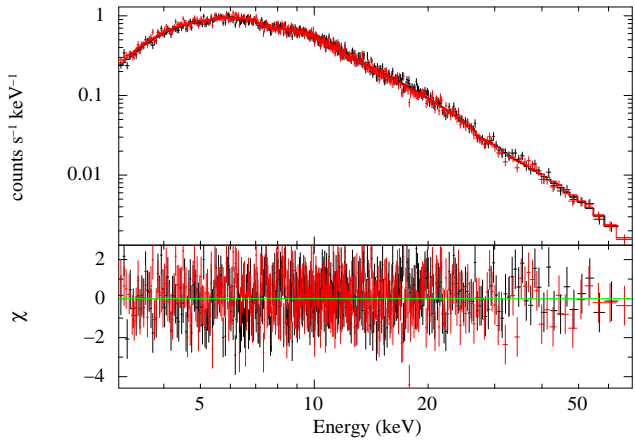}
\vspace{1cm}
\caption{\nus~spectra for epoch~2, fitted by a model with a PL and a multicolor disk blackbody. 
Spectra from FPMA and FPMB are in black and red, respectively. \label{fig_nuspec}}
\end{figure}

The low energy part of the spectrum, when fitted by a single continuum component shows
positive residuals below $\sim3.5$~keV. Therefore, 
in order to improve the modeling, we added a multicolor disk blackbody or {\it diskbb}
\citep{mit84,mak86} component to both
the above spectral models. 
Due to the limited soft X-ray coverage, we fixed the value of the inner disk temperature
as 0.3~keV, consistent with previous observations. The resulting 
$\chi^2/\nu$ is improved for both epochs. 
Assuming a distance of 8.5~kpc and inclination angles between
$30^{\circ}$ and $75^{\circ}$, the flux of the disk component points to values of the 
inner disk radius $\lesssim50$~$R_{\rm g}$, where $R_{\rm g}=GM/c^{2}$ is the 
gravitational radius (14.8~km for a 10${M}_{\rm \odot}$ black hole).
Figure \ref{fig_nuspec} shows the count rate spectrum, fitted by the PL plus low energy 
disk component for epoch~2. 

We note that the detection of the disk component in the hard X-ray spectra of black holes is 
common \citep{rei10}. In our case, it is quite unlikely that it can be ascribed to a calibration 
issue affecting the low energy response, which has been calibrated in flight against 
observations of the Crab source, known to have a PL spectrum below 10 keV. Also, the 
cross-calibration of \nus\/ with other X-ray instruments in the soft band is found to be very good
\citep{har13}. 

To better establish the presence of the soft excess, we fitted simultaneously 
the four \nus~spectra of the epochs 1 and 2 (note that the results of the spectral analysis 
for the two observations are in good agreement with each other). An F-test against the 
null hypothesis of a PL spectrum gives a very low probability of 3.6$\times10^{-4}$. 
However, following \citet{pro02}, 
the null distribution of the F-statistics could deviate from the nominal values    
when testing for the presence of a spectral feature, or the significance of an 
added model component. An alternative approach is using the Akaike information criterion, 
AIC \citep{aka74}. This method can be used to compare the relative quality of models, all 
using a given data set. Following \citet{bur02},  
for each model considered, {\it AIC} is computed as 
\begin{equation}
AIC = 2p-ln\left(L\right)+\frac{2p\left(p+1\right)}{n-p-1}
\label{eq:aic}
\end{equation}
where {\it p} is the number of free parameters, {\it n} is the number of data samples 
and {\it L} is the maximized value of the likelihood function applied 
to that model. In the case of $\chi^{2}$ fitting, 
\begin{equation}
ln(L) = C-\chi^{2}/2
\label{eq:constant}
\end{equation}
where {\it C} is a value, 
depending only on the data set and hence, is 
constant for all models. The model $\rm {\it M}_{k}$ which yields the minimum {\it AIC} is,
regardless of the true (and unknown) underlying process, the one minimizing the 
information loss and the relative likelihood of a model $\rm {\it M}_{i}$ can be estimated 
as $\exp$($\rm \Delta_{\it AIC}$/2), where $\rm \Delta_{\it AIC}$ is the difference in the {\it AIC} 
values of $\rm {\it M}_{i}$ and $\rm {\it M}_{k}$. 
Applying the above method to the same model spectra used in the F-test and computing the relative 
probability, we found that the PL model without the soft component is 3.8$\times10^{-3}$ less probable
than the corresponding model with the added {\it diskbb}. 
This result provides evidence that the disk 
blackbody is indeed present in our data. 
Using the same data set, we computed the upper limits for the equivalent width (EW) of an iron line 
feature at 6.4~keV. Modeling with a narrow line yields a 90\% confidence limit of 12~eV while a Gaussian 
line with a FWHM of 1~keV yields a corresponding value of 38~eV. These are statistical
limits and do not include an analysis of the possible fluctuations of the line background.

\begin{figure}
\includegraphics[angle=270,scale=.32]{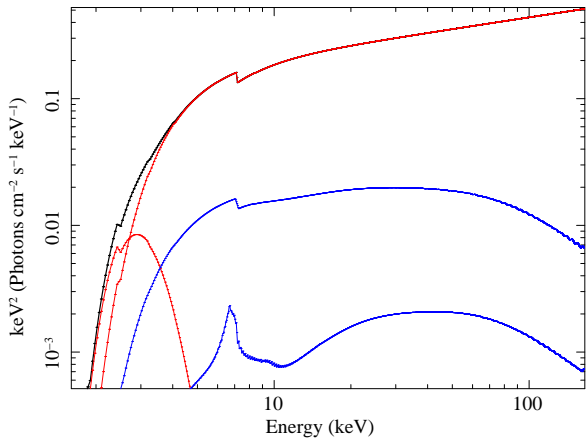}
\caption{Unfolded model spectrum of the {\it TBabs*(diskbb+powerlaw+reflionx)} fit for epoch~2 in black (total) and red
(components). There is no detection of the reflected component for this observation. The blue curves represent 
the 90\% upper limit reflected spectra using $\xi=1000$~erg~cm~s$^{-1}$ (lower curve) and 
$\xi=10,000$~ erg~cm~s$^{-1}$ (upper curve). 
\label{fig_reflionx}}
\end{figure}

We also used Comptonization models such as {\it comptt} \citep{tit94,hua95} and {\it compps} 
\citep{pou96} to fit the epoch~2 data. The latter includes modeling of reflection from the cold 
disk following the method of \citet{mag95}. In fitting with both models, we assume a Maxwellian 
distribution of electron temperature and a spherical geometry for the electron cloud. Moreover, 
for {\it compps} we used a disk viewing angle fixed at $60^{\circ}$.
A good fit was obtained in both cases when adding the {\it diskbb} component, i.e. $\chi^2/\nu$=753/740 and 
$\chi^2/\nu$=754/737 for {\it comptt} and {\it compps}, respectively. 
The fit with {\it comptt} sets a lower limit to the electron temperature of the hot, optically thin 
cloud as $kT_{\rm e}>$40~keV, whereas {\it compps} puts some more contraints on this parameter, yielding  
$kT_{\rm e}$=115$^{+41}_{-61}$~keV (90\% confidence). For the latter model a reflection 
component cannot be detected
and the 90\% upper limit to the reflection normalization is $\approx0.045$ for $\xi=$1000 erg~cm~s$^{-1}$,
where $\xi$ is the ionization parameter.
We also computed upper limits using {\it reflionx}, 
a recently improved version of the constant density, ionized disc model of \cite{ros99} 
and \cite{ros05}.  
For the spectral fitting, we tied the spectral index parameter of the {\it reflionx} component to 
the one of the direct PL component.  
Figure \ref{fig_reflionx} shows the best fit model with the {\it diskbb} and PL components
along with the 90\% upper limit spectra of the reflected components for a partially ionized 
and fully ionized disk ($\xi$=1000 erg~cm~s$^{-1}$ and $\xi$=10,000 erg~cm~s$^{-1}$, respectively). 
The corresponding 
upper limits in the solid angle normalization, ${\it R}=\Omega$/2$\pi$, for the above cases are
{\it R}=0.007 ($\xi$=1000~erg~cm~s$^{-1}$) and {\it R}=0.08 ($\xi$=10,000~erg~cm~s$^{-1}$). \\ 

\begin{table*}
\begin{center}
\caption{Spectral analysis of individual \nus~ observations \label{table_spec_nuonly} }
\begin{tabular}{lcllcccllc}
\tableline\\ [-2.0ex]
Model & Epoch & $N_{\rm H}$ & $\rm \Gamma$ & $E_{\rm fold}$ & $E_{\rm cut}$ & $N_{\rm disk}\times10^{-3}$  &  
$Fl_{\rm 3-10}$ & $Fl_{\rm 20-50}$ & $\chi^2/\nu$ \\
&       & ($10^{22}$~cm$^{-2}$) &              & (keV) & (keV)    &     &     &  \\
\tableline\\ [-2.0ex]
{\tt PL}          & 1 & 19.4$\pm$1.1         & 1.65$\pm$0.03 &        &             &                   &            
                343$\pm$9           & 478$\pm$12       & 408/377       \\
{\tt diskbb+PL}   & 1 & 24.3$^{+2.4}_{-2.2}$ & 1.71$\pm$0.35 &        &             & 144$^{+88}_{-65}$ &
                371$\pm$16          & 467$\pm$13       & 387/376       \\
{\tt PL}          & 2 & 19.4$\pm$0.7         & 1.67$\pm$0.02 &        &             &                   & 
                329$\pm$6           & 449$\pm$7        & 771/741       \\
{\tt diskbb+PL}   & 2  & 22.0$\pm$1.3         & 1.70$\pm$0.02 &       &              & 66$^{+36}_{-30}$  &
                343$\pm$8           & 444$\pm$5        & 755/740       \\
{\tt CPL}         & 2  & 18.3$\pm$0.9         & 1.58$\pm$0.04 & 215$^{+232}_{-74}$ &        &           & 
                320$\pm$7           & 442$\pm$8        & 761/740       \\
{\tt highecut*PL} & 2 & 18.2$^{+0.9}_{-1.2}$  & 1.60$^{+0.03}_{-0.07}$ & 260$^{+268}_{-122}$ & $<14.4$ & & 
                319$\pm$9           & 441$\pm$8        & 760/739       \\
{\tt diskbb+CPL}  & 2  & 20.9$^{+1.5}_{-1.8}$ & 1.65$^{+0.02}_{-0.06}$ & $>160$ &    & 49$\pm$31         & 
                335$^{+6}_{-12}$    & 441$\pm$8        & 753/739       \\
\tableline\\ [-2.0ex]
\end{tabular}
\tablecomments{$N_{\rm H}$ is the hydrogen absorption column estimated with the model {\it TBabs}, 
$\Gamma$ the PL photon index, $E_{\rm fold}$ the e-folding energy, $E_{\rm cut}$ the cutoff
energy of the {\em highecut} model, and $N_{\rm disk}$ is the normalization of the {\em diskbb} model.
$Fl_{\rm 3-10}$ and $Fl_{\rm 20-50}$ are the flux values measured of the unabsorbed emission of the 
PL or CPL components, in the 3-10~keV and 20-50~keV bands, and are given in units of 
$10^{-12}$~erg~s$^{-1}$~cm$^{-2}$. All errors are computed as 90\% confidence. 
} 
\end{center}
\end{table*}

\begin{figure}
\includegraphics[angle=0,scale=.47]{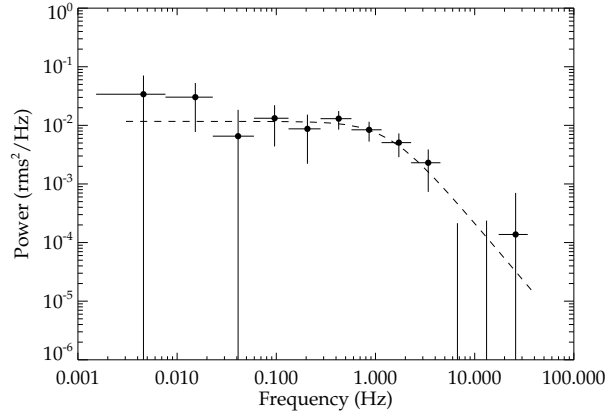}
\caption{\nus~power spectrum measured in epoch~2 and averaged for both FPMs.  See text for details. 
\label{fig_powspec}}
\end{figure}

{\em Timing analysis ---} 
We extracted light curves of \targ~ in the energy range 3--60~keV from epoch~2 with XSELECT~V2.4b using 
a time resolution of 10ms, and corrected them using the tool {\it nulccorr}. Prior to temporal binning, 
the filtered event files were corrected for arrival times at the Solar System Barycenter using the JPL 2000 ephemeris 
(for this purpose, we used the {\it barycorr} tool in the HEASOFT v13 distribution). We then calculated 
power spectra on different contiguous sections of the light curves and averaged them into a 
total spectrum. Each single power spectrum was built using intervals of 32768 bins and averaging 
up to 10 intervals in a frame. The total spectrum was finally rebinned in frequency channels for more statistics.
An offset constant term was subtracted from the total spectrum to remove the Poisson noise level and compensate for 
residual effects of the deadtime correction.
This term was evaluated as the average power in the frequency interval 10--49~Hz.   
The power spectrum is shown in Figure \ref{fig_powspec}. Although the statistics are quite poor due to the 
short exposure time of the observations, we could model the power spectrum with a zero-centered Lorentzian 
function and obtained a 
good fit ($\chi^2/\nu$=15.4/23). For our adopted normalization, the root-mean-square (RMS) variability is derived as 
the integral of the Lorentzian. This resulted in $15.4\pm2.2$\%. The error also includes the uncertainty in the
evaluation of the offset term.

\subsection{\intg} \label{analysis_integral}
We extracted images, spectra and lightcurves of \targ~ from the IBIS/ISGRI instrument using the 
\intg~Off Line Analysis v.10 package
which takes advantage of a recently updated calibration \citep{cab12}. The lightcurves and
spectra of \targ~ were extracted by simultaneous fitting of all the sources detected in the mosaic 
images above a detection level of $\approx$~7$\sigma$. This process reduces systematic 
noise in the reconstructed flux due to the cross-talk between sources to a level close to $\sim1$\% or
less. For \targ~ this effect is mostly limited to lower energies, because the source is the
brightest one in the field at $E>50$~keV. For this reason we excluded the energy channels below 26~keV from 
the analysis. 

\begin{figure}
\centering
\includegraphics[angle=0,scale=0.50]{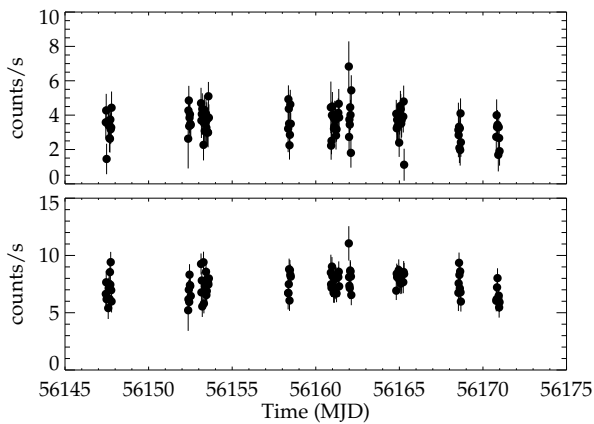}
\caption{Light curves measured by IBIS/ISGRI for epoch~3, in the energy intervals 26--60 keV (bottom) and
60--150 keV (top).
\label{fig_ibislc}}
\end{figure}

\begin{figure}
\includegraphics[angle=270,scale=.32]{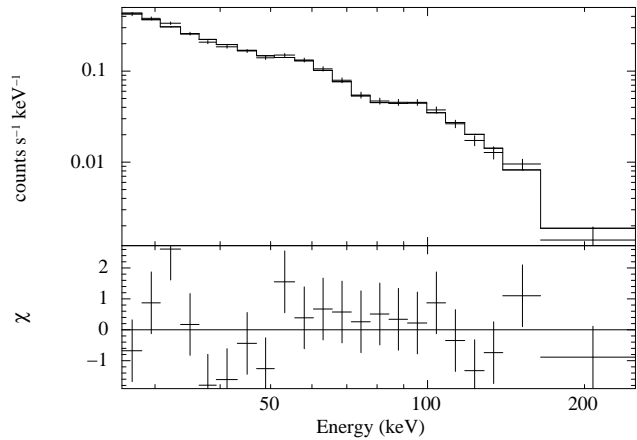}
\vspace{1cm}
\caption{Total energy spectrum of IBIS/ISGRI modeled by a PL with high energy cutoff. 
\label{fig_spec_ibis}}
\end{figure}

Figure \ref{fig_ibislc} shows the IBIS/ISGRI light curves in the energy bands 26--60~keV and
60--150~keV for epoch~3. We performed an analysis
of the IBIS spectrum to search for any evidence of spectral
curvature at high energy, which is a typical signature of (thermal) Comptonization. The absorption value was
fixed as $N_{\rm H}$=$2\times10^{23}$~cm$^{-2}$ in {\em TBabs} and a systematic error of 1\% has been added
to all spectral channels. For IBIS, the PL model fit is not satisfactory, resulting in $\chi^2/\nu$=48.1/20. 
Conversely, using the CPL model yields $\chi^2/\nu$=24.2/19 with a high energy cutoff at
$E_{\rm fold}$=123$^{+69}_{-34}$~keV. The resulting count rate spectrum with the convolved best 
fit model is shown in Figure \ref{fig_spec_ibis}. This measurement yields definitive evidence for spectral 
curvature at high energies and is fully compatible with the \nus~measurements 
described in Sect. \ref{analysis_nustar}. See also the following Sect.\ref{analysis_all}. 

\subsection{ {\em NuSTAR}/{\em INTEGRAL} spectrum} \label{analysis_all}

\begin{table*}
\begin{center}
\caption{Model parameters for the \nus/\intg~ spectrum \label{table_all5} }
\scriptsize
\begin{tabular}{lccccccccccc}
\tableline\\ [-2.0ex]
Model & $N_{\rm H}$ & $\rm \Gamma$ & $E_{\rm fold}$ & $kT_{\rm e}$ & $\tau$ & $N_{\rm disk}$ & $kT_{\rm in}$ & $C_{\rm 3-2}$ & 
  $Fl_{\rm 20-50}$ & $Fl_{\rm 50-200}$ & $\chi^2/\nu$ \\
      & ($10^{22}$~cm$^{-2}$) &    & (keV)         & (keV)        &        & $\times10^{-3}$             & (keV)         &  & 
                   &                   &              \\
  \tableline\\ [-2.0ex]

{\tt PL}              & 20.4$\pm$0.6 & 1.690$\pm$0.013 & & & & & & 1.04$\pm$0.03 &  
   441$\pm$6 & 955$\pm$26 & 1337/1141 \\

{\tt diskbb+PL}       & 27.2$\pm$1.2 & 1.758$\pm$0.021 & & &           & 14$^{+51}_{-9}$  & 0.38$\pm$0.06 & 1.10$\pm$0.03 &  
   432$\pm$6 & 866$\pm$32 & 1262/1139 \\

{\tt CPL}        & 18.0$\pm$0.6 & 1.556$\pm$0.024 & 163$^{+29}_{-22}$ & & & &                                & 1.27$\pm$0.03 & 
   450$\pm$6 & 623$\pm$29 & 1188/1140 \\

{\tt diskbb+CPL} & 20.3$\pm$1.3 & 1.597$\pm$0.029 & 190$^{+42}_{-30}$ & & & 47$^{+27}_{-22}$ & 0.3$^{(f)}$        & 1.13$\pm$0.03 &
   438$\pm$6 & 625$\pm$28 & 1174/1139 \\

{\tt diskbb+comptt}   & 20.4$\pm$0.9         &  &  & 38$^{+9}_{-5}$ & 1.41$^{+0.17}_{-0.26}$ & 48$^{+23}_{-20}$ & 0.3$^{(f)}$       & 1.13$\pm$0.03 &  
   441$\pm$7 & 638$\pm$31 & 1168/1139 \\
\\
     \tableline\\ [-2.0ex]
\end{tabular}
\footnotesize
\tablecomments{
$kT_{\rm in}$ is the inner disk temperature and $C_{\rm 3-2}$ is the multiplicative normalization constant of epoch~3 versus epoch~2 observations. See Table \ref{table_spec_nuonly} for description of the other model parameters. 
All errors are computed as 90\% confidence. 
}
\end{center}
\end{table*}

We fitted the \nus~ and IBIS/ISGRI data for the whole set of observations described in Table 
\ref{table_observations}. For the spectral fits we considered, in addition to the empirical PL and 
CPL models, the thermal Comptonization model {\it comptt}, which assumes that the soft seed
photons in the accretion disk are diffused via inverse Compton scattering by  
an electron plasma with a Maxwellian distribution of energies. The spectrum 
of the seed photons is assumed to follow the Wien law and is typically $kT<0.5$~keV. For our purposes, 
the shape of the seed photon spectrum is not relevant since our spectra are measured at energies 
$>3$~keV. Conversely, the high energy part of the spectrum is dominated by the cutoff induced by the
finite temperature of the plasma electrons. 

\begin{figure}
\includegraphics[angle=270,scale=.32]{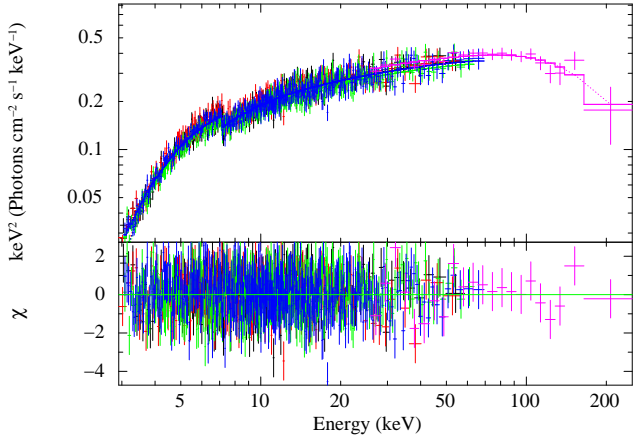}
\vspace{0.8cm}
\caption{Unfolded total spectrum of the 1E1740.7-2942 low/hard state, folded through the {\it diskbb}+{\it comptt} model.
Black (FPMA) and red (FPMB) spectra are for epoch~1, green (FPMA) and dark blue (FPMB) for epoch~2, magenta for 
epoch~3 (IBIS).  
\label{fig_results_all5}}
\end{figure}

The data set consists of four \nus~ spectra and one IBIS/ISGRI spectrum. We allowed free normalization
between the three different observations by including a normalization factor in all the models. The \nus~ spectra
from FPMA and FPMB of the same observation are cross-calibrated between each other at the few \% level. For 
this reason, we included another free normalization factor. The five spectra are the same discussed previously in this
Section. We summarize the results of the overall fits in Table~\ref{table_all5}. In Figure 
\ref{fig_results_all5}, the resulting unfolded spectrum for the {\it diskbb+comptt} model is shown. The
result of the fit, even if not formally acceptable, is adequate to describe the broadband shape 
and characterize it as dominated by a thermal Comptonized component in the energy band 3-250 keV. Adding 
the soft component improves the quality of all model fits; however, only for one model ({\it diskbb+PL})
it is possible to constrain both disk temperature and normalization.    

Table \ref{table_all5} also lists the values of the relative normalization factor $C_{3-2}$ between 
the IBIS/ISGRI and the \nus~epoch~2 observations. In the fit, we fixed to unity the normalization of the epoch~2 
observation for the \nus~FPMA. The relative (variable) normalization of epochs 1 to 2 
is actually found to be constant for all model fits, i.e. $C_{1-2}$=1.04$\pm$0.01.  
Note that the value of $C_{3-2}$ accounts not only for the different flux between the two observations
but also for any difference in the cross-calibration of the two instruments. 

\begin{figure}
\includegraphics[angle=270,scale=.32]{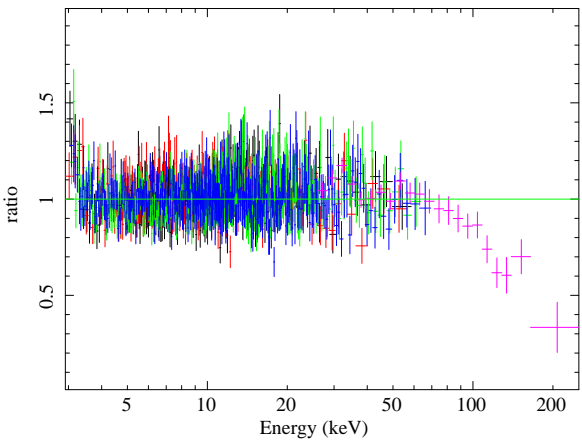}
\vspace{1cm}
\caption{Data/model ratio from joint fitting through a PL of the \nus~ and \intg~ data. See Fig.\ref{fig_results_all5}
for color coding of the spectra.
\label{fig_ratios}}
\end{figure}

Finally, to better describe the departure of the spectrum from a pure PL we plot in Figure~\ref{fig_ratios} the 
data/model ratios in the energy band 3--250~keV. The positive residuals at low energies are likely due to the 
thermal disk component, while the negative values of the ratio at high energies are due to an exponential 
cutoff.   

\section{Discussion} \label{sect_disc}

These observations show that \targ~presents a featureless spectrum in its hard state and is 
consistent with being dominated by Comptonization with an electron temperature of $\sim40$~keV. 
Assuming a distance of 8.5~kpc, the luminosity of the source is $\approx2.2\times10^{37}$~erg~s$^{-1}$, i.e. a fraction
less than a few \% of the Eddington limit for a stellar black hole.
The sensitivity of \nus~constrains the strength of the reflection   
component to be quite low, similar to the limits which have been measured previously 
by \citet{rey10}, who reported a disk reflection fraction $\lesssim0.1$ at the 99\% confidence level. 
However, in contrast to these observations, the \suz~data are compatible with no high energy cutoff
($E_{\rm fold}\gtrsim350$~keV). This 
issue could be ascribed to either a genuine variation of the plasma temperature causing the cutoff
energy to vary, to a different inherent mechanism for the hard X-ray emission,
or to a difference in calibration with the HXD instrument.
However, we note that the \suz~and \nus~PL slopes are similar when data are fitted without a high 
energy cutoff. Also, a sensitivity bias is unlikely as   
\citet{rey10} report a \suz~detection up to 300~keV. Therefore, it is likely that the difference is related 
to a genuine variation of the physical conditions of the high energy corona in two different occurrences
of the hard state. Both \suz~and \nus~detect a soft component in the 
spectrum of the hard state and the \nus~data are compatible with a disk inner radius close to the ISCO.
However, we cannot exclude a recessed disk as we cannot determine any reliable constraints of the 
parameters of the soft component. Using {\em diskbb} on the single \nus~observations and 
leaving $kT_{\rm in}$ free, the fitting procedure converges to either values $\lesssim0.15$~keV with a 
rather strong soft component (rather unlikely in a hard state), or to values $kT_{\rm in}$~$\sim0.3-0.4$~keV, 
yielding normalizations similar to what was observed by \suz. \citet{cas12} also report the presence of 
a soft component in the hard state during an observation in 2005 by {\em XMM/Newton}, with an inner disk 
color temperature close to 0.25~keV. Our upper limits on the EW of a 6.4~keV Fe~line are consistent with
the ones reported by \citet{nak10}, who found 8~eV with \suz/XIS, and by \citet{sak99}, who 
measured EW$<15$~eV with ASCA. \citet{hei98} also report a line with EW=19$^{+19}_{-14}$~eV with observations
by {\em RXTE}/PCA.

At the highest energies we can compare our observations with the available 
measurements from \intg/SPI. These observations, described in \citet{bou09}, point to the presence of 
two different spectral components, of which the lower in energy is well described by thermal 
Comptonization with a cutoff energy of $\sim$140~keV. The higher energy component is above 
$\sim$200~keV. We did not detect this component: fitting   
our \nus/\intg~spectrum with a two temperature model,
in which the seed photon temperature of the high energy component is equal to 30~keV as in \citet{bou09},
the 90\% upper limit flux for the high energy excess is 4$\times10^{-10}$~erg~s$^{-1}$~cm$^{-2}$ in the 
20--250~keV band. This value, scaled by the intensity measured by SPI, is $\sim20$\% higher than the 
flux of this component reported by these authors. Our observations are then substantially
in agreement with those of \citet{bou09} for the highest energies. 
 
There are several possible reasons for the low level of reflection from \targ. If the accretion disk is close
to the ISCO and the upper layers of the disk are close to being fully ionized, the reflection would 
be non-negligible but the shape of the reflected spectrum could be quite similar to that of the 
directly radiated flux (see e.g. Figure~\ref{fig_reflionx}). In this case, it would be quite difficult 
to detect it. Furthermore, Fe lines and edges 
can be significantly broadened by additional Doppler and relativistic effects. Conversely, if the disk is 
recessed, the geometry and positioning of the corona relative to the disk could play 
a role and likewise for the (unknown) disk inclination. A low level of reflection is also expected if a 
substantial part of the hard X-ray spectrum is generated in an outflow \citep{bel99}. However, the 
mechanism of acceleration in an outflow or jet and its connection to the hot disk corona are  
presently not understood and so, also the relative contribution of the jet 
component. The hard tail detected by \citet{bou09} could be non-thermal and
associated with a jet, similar to the high energy polarized component in Cyg X-1. In this case 
we could have a scenario in which the bulk of hard X-rays are indeed produced by a hot corona up to 
$\sim200$~keV, with a jet dominating emission at higher energies. 

For \targ~we report clear evidence for a 
high energy cutoff, which is a typical signature of a thermal electron plasma. 
However, this scenario is not the only one that is consistent with such a cutoff.
The emergent spectra computed by jet models could also contain high energy exponential cutoffs in 
the region from $\sim$100 to a few hundred keV, generated, e.g., by synchrotron cooling as described in 
\citet{mar01}. These features can also be produced by Comptonized emission from regions within the jet 
\citep{rei03,gia05} or at the base of it \citep{mar05}. For the jet model of \citet{mar05}, the interplay 
of the direct synchrotron and Comptonized components controls the hardening of the spectrum above 
10 keV, which is commonly ascribed to reflection in a corona/disk model. If such hardening is not present, 
as in the case of \targ, the synchrotron emission component could be dominant. However,
we emphasize that in our case it is not possible to provide any direct evidence of synchrotron emission 
or non-thermal Comptonization, and our broadband spectrum in the range 3--250 keV is well described by 
inverse Compton radiation by a hot, optically thin thermal plasma plus a soft disk component.        
 
\section{Conclusions} \label{sect_conc}

We have analyzed spectra of the well known microquasar \targ, located in the vicinity of the 
GC using the \nus~ telescope and the hard X-ray instrument IBIS/ISGRI on board \intg.
During the observations, the source was in a typical low/hard state. 
We have analyzed \nus~spectra from 
two different observations (taken about two weeks apart) and found that they are fully 
consistent. The \nus~and IBIS/ISGRI data, spanning a time range of $\approx1.5$~months, 
are also well in agreement both for the spectral modeling and for the relative normalization. 
The broadband spectrum in the range 3--250~keV is essentially modeled by a component 
that is consistent with Comptonization with a thermal energy cutoff $kT_{\rm e}\sim40$~keV.
The \nus~power spectrum is compatible with what was previously observed 
by \xte~\citep{smi97,lin00} and also RMS variation is detected at the level of $\sim15$\%. 

At the softest energies, near 3~keV, there is some evidence for 
a soft component. Although we cannot obtain a reliable measure of the disk temperature and inner radius,
due to both the low threshold of 3~keV and the shortness of the \nus~observations, 
the disk component observed is quite compatible with previous observations 
by \suz. Conversely, the detection of a high energy cutoff points to a possible change in the 
physical conditions of the plasma in the Comptonizing corona (we also note that a state change occurred
shortly after these observations). 

The very high sensitivity of \nus~has allowed us to characterize in detail the spectrum
of this source up to $\sim$70~keV and combining \nus~and \intg~provides a determination of the
overall properties of the broadband emission. 
The complementarity of \nus~and \intg~is 
excellent for the study of the properties of black holes in the low/hard state,
for which the bulk of the emission is in the energy range $\sim50-120$~keV.  
In the case of \targ, we were able to rule out a single PL model for the broadband 
spectrum. \nus~could not detect any reflection feature; 
however, more interesting results/constraints will likely come from significantly 
longer observations of this source.  

\acknowledgments
This work was supported under NASA Contract No.
NNG08FD60C, and made use of data from the \nus~mission, 
a project led by the California Institute of Technology,
managed by the Jet Propulsion Laboratory, and
funded by the National Aeronautics and Space
Administration. We thank the \nus~Operations, Software
and Calibration teams for support with the execution
and analysis of these observations. This research has
made use of the \nus~Data Analysis Software 
({\em NuSTARDAS}) jointly developed by the ASI Science Data
Center (ASDC, Italy) and the California Institute of
Technology (USA). LN wishes to acknowledge
the Italian Space Agency (ASI) for financial support 
by ASI/INAF grants I/037/12/0-011/13 and I/033/10/0 
and the engineering support of M.Federici for setup and 
maintenance of the \intg~archive and Data Analysis 
Software at IAPS. MB wishes to acknowledge the support 
from the Centre National d'Etudes Spatiales (CNES).

\end{document}